\documentclass[peerreview,draftcls,12pt,onecolumn]{IEEEtran}

\usepackage{graphicx}
\usepackage{psfrag}
\usepackage[dvips]{color}
\usepackage{bm}
\usepackage{amssymb,amsmath}
\usepackage{multirow}
\usepackage{subfigure}
\usepackage{algorithm}
\usepackage{algorithmicx}
\usepackage{algpseudocode}
\usepackage{soul}

\usepackage{dsfont}

\usepackage[normalem]{ulem}
\usepackage[user]{zref}

\newcounter{revc}
\makeatletter \zref@newprop{revcontent}{} \zref@addprop{main}{revcontent}
\zref@newprop{revsec}{} \zref@addprop{main}{revsec}

\newcommand{\revi}[2]{%
\zref@setcurrent{revsec}{\thesection}%
\zref@setcurrent{revcontent}{#2}%
\refstepcounter{revc}%
\label{#1}
\zlabel{#1}%
\uline{#2} }

\newcommand{\revr}[2]{%
\zref@setcurrent{revsec}{\thesection}%
\zref@setcurrent{revcontent}{#2}%
\refstepcounter{revc}%
\zlabel{#1}%
\label{#1} 
\sout{#2}
} \makeatother

\newcommand{\argmax}{\operatornamewithlimits{argmax}}

\IEEEoverridecommandlockouts

\title{A Dynamic Clustering and Resource Allocation Algorithm for Downlink CoMP Systems with Multiple Antenna UEs}

\author{P.~Baracca,
~F.~Boccardi, 
and~N.~Benvenuto
\thanks{Part of this work has been presented at the International Symposium on Wireless Communication Systems (ISWCS) 2012, Paris (France), and at the International Conference on Signal Processing, Computing and Control (ISPCC) 2013, Shimla (India).}%
\thanks{P. Baracca and F. Boccardi are with Bell Laboratories, Alcatel-Lucent (email: \{paolo.baracca, federico.boccardi\}@alcatel-lucent.com).}%
\thanks{N. Benvenuto is with the Department of Information Engineering, University of Padova, Italy (email: nb@dei.unipd.it).}%
}

\begin{document}

\setcounter{page}{1}

\maketitle

\begin{abstract}
Coordinated multi-point (CoMP) schemes have been widely studied in the recent years to tackle the inter-cell interference. In practice, latency and throughput constraints on the backhaul allow the organization of only small clusters of base stations (BSs) where joint processing (JP) can be implemented. In this work we focus on downlink CoMP-JP with multiple antenna user equipments (UEs) and propose a novel dynamic clustering algorithm. The additional degrees of freedom at the UE can be used to suppress the residual interference by using an interference rejection combiner (IRC) and allow a multi-stream transmission. In our proposal we first define a set of candidate clusters depending on long-term channel conditions. Then, in each time block, we develop a resource allocation scheme by jointly optimizing transmitter and receiver where: a) within each candidate cluster a weighted sum rate is estimated and then b) a set of clusters is scheduled in order to maximize the system weighted sum rate. Numerical results show that much higher rates are achieved when UEs are equipped with multiple antennas. Moreover, as this performance improvement is mainly due to the IRC, the gain achieved by the proposed approach with respect to the non-cooperative scheme decreases by increasing the number of UE antennas.
\end{abstract}



\section{Introduction}

Coordination among base stations (BSs) has been widely studied in the recent years to tackle inter-cell interference which strongly limits the rates achieved in cellular systems, in particular by the user equipments (UEs) at the cell-edge \cite{comp_book}. Supported by the first results promising huge gains with respect to the baseline non-cooperative system \cite{karakayali_aug06}, a lot of attention has been paid to the topic both in the academia \cite{comp_tutorial, bjornson_book} and in the industry \cite{3gpp_comp, hp_mayer}. These techniques, known in the industry as coordinated multi-point (CoMP), are classified into a) coordinated scheduling/beamforming (CS/CB), which require channel state information (CSI) but no data sharing among the BSs, and b) joint processing (JP), which require both CSI and data sharing among the BSs. This paper focuses on downlink CoMP-JP, where BSs jointly serve the scheduled UEs by sharing the data to be sent. Although CoMP-JP is a very promising technique, many issues make its implementation still challenging. First, CSI at the transmitter may be unreliable because of noise on channel estimation in time division duplex (TDD) systems and limited bandwidth available for feedback in frequency division duplex (FDD) systems. Then, sharing UE data among all the BSs is generally limited by throughput and delay constraints in the backhaul infrastructure. A possible approach to deal with backhaul throughput constraints relies on partial sharing of UE data among the BSs, i.e., a BS serving a certain UE may have only a partial knowledge of the data to be sent toward that UE \cite{zakhour_tsp11,baracca_const_quantization}. Although the promising results achieved under idealistic assumptions, partial UE data sharing has not found application in a real system mainly because of its complexity. Hence, most of the works in the literature focus on the simpler clustering approach to deal with limited throughput backhaul, where the BSs are organized in clusters and joint processing is applied within each cluster by sharing the whole data to be sent among all the BSs of the cluster. However, even if intra-cluster interference is mitigated by using CoMP schemes within the cluster, UEs at the cluster border suffer strong inter-cluster interference (ICI).
Many clustering schemes have been developed in the literature to deal with ICI. In \cite{heath_tcom09} static clustering with block diagonalization is considered and precoders are designed in each cluster by nullifying the interference towards UEs of neighboring clusters close to the border. A more flexible solution is obtained with dynamic clustering \cite{papadogiannis_icc08,boccardi_asilomar08} where the set of clusters changes over time by adapting to the network conditions. In \cite{papadogiannis_icc08} a greedy algorithm is developed where, for each cluster, the first BS is selected randomly to guarantee fairness, while the remaining BSs of the clusters are selected by maximizing the cluster sum rate. In \cite{boccardi_asilomar08} a joint clustering and scheduling scheme is developed by generating a set of candidate clusterings based on the long-term channel conditions and then by selecting, in each time block, the best one. In \cite{moon_vtc11} the set of clusters is optimized by maximizing the increase in the achievable UE rate, whereas in \cite{moon_vtc11,liu_wcsp09} by minimizing the interference power. In \cite{zhou_globecom09} a BS negotiation algorithm is designed for cluster formation by considering a fixed cluster size. In \cite{weber_vtcspring11} active clusters are selected by minimizing an overall cost function among a set of candidates which depend on UE average received power.  A framework for feedback and backhaul reduction is developed in \cite{papadogiannis_jan11} where each UE feeds back CSI only to a subset of BSs, and UEs associated to the same subset are grouped together. In \cite{gong_globecom11} a greedy scheduling algorithm with overlapping clusters is proposed where precoders are designed by considering the layered virtual signal to interference plus noise ratio (SINR) criterion \cite{zakhour_aug10}.

However, most of the works on dynamic clustering (\cite{papadogiannis_icc08, boccardi_asilomar08, moon_vtc11, liu_wcsp09, papadogiannis_jan11, gong_globecom11}) assume that UEs are equipped with only one antenna, although the Long Term Evolution (LTE) Advanced standard developed by the 3rd Generation Partnership Project (3GPP) considers that UEs may be equipped with up to eight antennas \cite{boccardi_2012}. Although this number seems a bit optimistic for current mobile devices, the technological innovation may allow in the near-future manufacturing smartphones or tablets with numerous antennas and hence much more attention should be paid to the study of CoMP schemes with multiple antenna UEs \cite{hwang-feb13, clercks_apr13}. Therefore, in this work we consider downlink CoMP-JP with a constraint on the maximum cluster size and propose a novel dynamic clustering algorithm by explicitly considering that UEs are equipped with multiple antennas. In our proposal, UEs can exploit these additional degrees of freedom by implementing interference rejection combiner (IRC) \cite{irc_combiner_winters} to partially suppress the ICI and being served by means of a multi-stream transmission. Moreover, differently from many works on dynamic clustering where UE selection is not considered and a simple round robin scheduler is implemented (\cite{papadogiannis_icc08, moon_vtc11, liu_wcsp09, zhou_globecom09, papadogiannis_jan11}), here we assume UE scheduling as a part of the optimization. In our approach we first define a set of candidate clusters depending on long-term properties of the channels. Then, given this set of candidate clusters, in each time block the proposed algorithm follows a two-step procedure: a) a weighted sum rate is estimated within each candidate cluster by performing UE selection, precoding design, power allocation and transmission rank selection, and then b) the central unit (CU) coordinating all the BSs schedules a set of non-overlapping candidate clusters by maximizing the system weighted sum rate. We emphasize that the developed resource allocation is performed by jointly optimizing transmitter and receiver, i.e., by taking into account the benefits of IRC and multi-stream transmission. Moreover, the proposed scheme can be implemented in a semi-distributed way, as the cluster weighted sum rate optimization can be performed within the cluster without additional information from the out-of-cluster BSs.

We evaluate the proposed solution in a LTE-TDD scenario, where channels are estimated at the BSs thanks to pilot sequences sent by the UEs, and compare it against a baseline single-cell processing (SCP) scheme and two static clustering schemes, where the clusters do not dynamically adapt to the network conditions. Numerical results with perfect CSI at the BSs show that the achievable rates strongly increase with the number of UE antennas. Then, as with CoMP part of the interference is managed at the transmit side, multi-stream transmission is more effective with the proposed scheme than with SCP. However, as most of the gain is due to the interference suppression capability of the IRC, the relative gain achieved by the proposed scheme with respect to SCP decreases by increasing the number of UE antennas. Finally, a further decrease of this gain is observed when imperfect CSI is considered at the BSs.

{\em Notation}: We use $\left( \cdot \right)^{T}$ to denote transpose and $\left( \cdot \right)^{H}$ conjugate transpose. $\bm{0}_{N \times M}$ denotes the matrix of size $N \times M$ with all zero entries, $\bm{I}_{N}$ the identity matrix of size $N$, ${\rm tr}(\bm{X})$ the trace of matrix $\bm{X}$, ${\rm det}(\bm{X})$ the determinant of matrix $\bm{X}$, ${\rm vec}(\bm{X})$ the vectorization of $\bm{X}$, $\left\|\bm{X}\right\|$ the Frobenius norm of $\bm{X}$, $\left[\bm{X}\right]_{n,m}$ the entry on row $n$ and column $m$ of $\bm{X}$, $\left[\bm{X}\right]_{\cdot,m}$ the $m$-th column of $\bm{X}$, and ${\rm diag}\left(\bm{x}\right)$ the diagonal matrix with the entries of vector $\bm{x}$ on the diagonal. Expectation is denoted by $\mathbb{E} \left[ \cdot \right]$.

\section{System Model}

We consider a TDD system where a set of BSs $\mathcal{J}=\{1,2,\ldots,J\}$, each equipped with $M$ antennas, is serving a set of UEs $\mathcal{K}=\{1,2,\ldots,K\}$, each equipped with $N$ antennas, with $K > JM$.  As the overall number of transmitting antennas is not sufficient to serve all the UEs in the same time and frequency band, UE scheduling is part of the optimization problem. We assume a block fading channel model and denote with $\bm{H}_{k,j}(t)$, $t=0,1,\ldots,T-1$, the multiple-input multiple-output (MIMO) channel matrix of size $N \times M$ between BS $j$ and UE $k$ in block $t$. We consider that the entries of matrix $\bm{H}_{k,j}(t)$ are identically distributed zero-mean complex Gaussian random variables, i.e., $\left[\bm{H}_{k,j}(t)\right]_{n,m} \sim \mathcal{CN}\left(0,\sigma_{k,j}^2\right)$, for $n=0,1,\ldots,N-1$ and $m=0,1,\ldots,M-1$, where $\sigma_{k,j}^2$ represents the large scale fading between BS $j$ and UE $k$, which depends on path-loss and shadowing. We assume that the statistical description of the channels does not change for all the $T$ blocks, whereas fast fading realizations are independent among different blocks. Then, we denote with $\bm{\Sigma}_{k,j} = \mathbb{E}\left[ {\rm vec}(\bm{H}_{k,j}(t)) {\rm vec}(\bm{H}_{k,j}(t))^{H} \right]$ the covariance matrix of the channel matrix $\bm{H}_{k,j}(t)$. We indicate with $N_E$ the number of resource elements, i.e., time slots, forming a block. Note that the block fading model considered in this work can be adapted to represent a more realistic channel which changes continuously both in time and in frequency domains by suitably selecting the number of resource elements in each block. In fact, by denoting with $W_C$ and $T_C$ the coherence bandwidth and time of the channel, respectively, we have $N_E = W_C T_C$.

We assume that the BSs are coordinated by a CU and the backhaul links have zero latency and are error free. Each block is organized in three phases: a) in the first phase all the UEs send pilot sequences to allow channel estimation at the BSs, b) in the second phase clustering, UE scheduling, beamforming design, transmission rank selection and power optimization are performed by the CU and finally c) in the third phase the BSs perform data transmission toward the set of scheduled UEs.

\subsection{First Phase: Uplink Pilot Transmission}

The first $N_T$ resource elements of each block are allocated to the uplink pilot transmission performed by the UEs. We assume that orthogonal sequences are employed by the UEs, thus interference on channel estimation is avoided at the BSs: in the considered scenario this is achieved by imposing a minimum length of the training sequence of $N_T \geq NK$. By denoting with $P^{({\rm UE})}$ the maximum power available at each UE and $\sigma_n^2$ the thermal noise power, under the assumption of perfect reciprocity BS $j$ estimates the channel matrix $\bm{H}_{k,j}(t)$ connecting UE $k$ to itself from the observation
\begin{equation}
\begin{split}
o_{k,j,n,m}(t) = & \left[\bm{H}_{k,j}(t)\right]_{n,m} + \eta_{k,j,n,m}(t)\,,\\
& n=0,1,\ldots,N-1,\,m=0,1,\ldots,M-1\,,
\end{split}
\label{obs_ce}
\end{equation}
where $\eta_{k,j,n,m}(t)\sim\mathcal{CN}\left(0,\frac{N\sigma_n^2}{N_T P^{({\rm UE})}}\right)$. By assuming that BS $j$ knows the covariance matrix $\bm{\Sigma}_{k,j}$, the minimum mean square error (MMSE) estimate $\hat{\bm{H}}_{k,j}(t)$ of $\bm{H}_{k,j}(t)$ given the observation (\ref{obs_ce}) can be written as \cite[Ch.~10]{Kay-estimation}
\begin{equation}
\begin{split}
{\rm vec}\left(\hat{\bm{H}}_{k,j}(t)\right) = & \bm{\Sigma}_{k,j} \left( \bm{\Sigma}_{k,j} + \frac{N\sigma_n^2}{N_T P^{({\rm UE})}} \bm{I}_{MN} \right)^{-1} \times \\
& \left( {\rm vec}\left(\bm{H}_{k,j}(t)\right) + {\rm vec}\left(\bm{\eta}_{k,j}(t)\right) \right)\,,
\end{split}
\label{ce_w_corr}
\end{equation}
where $\left[\bm{\eta}_{k,j}(t)\right]_{n,m} = \eta_{k,j,n,m}(t)$.

Note that in the case of uncorrelated channels, i.e., when $\bm{\Sigma}_{k,j} = \bm{I}_{MN}$, the expression in (\ref{ce_w_corr}) turns out to be
\begin{equation}
\hat{\bm{H}}_{k,j}(t) = \frac{1}{1 + \frac{N\sigma_n^2}{N_T P^{({\rm UE})} \sigma_{k,j}^2}} \left( \bm{H}_{k,j}(t) + \bm{\eta}_{k,j}(t) \right)\,.
\label{ce_wo_corr}
\end{equation}

\subsection{Second Phase: Resource Allocation at the CU}

After uplink pilot transmission, the CU organizes BSs in clusters and schedules in each block $t$ a subset $\mathcal{S}(t)\subseteq\mathcal{K}$ of UEs. In this work we consider dynamic multi-stream transmission and we denote with $l_k(t)$ the transmission rank allocated to UE $k$ in block $t$, i.e., the number of streams sent toward UE $k$. Let us denote with $\bm{G}_{k,j}(t)$ the $M\times l_k(t)$ beamforming matrix used by BS $j$ to serve UE $k$, $\bm{P}_k(t)=\left[ P_{k,0}(t) , P_{k,1}(t) , \ldots, P_{k,l_k(t)-1}(t) \right]$ the power allocation vector for UE $k$, and with $P^{({\rm BS})}$ the power available at each BS. Then, as the BSs do not have perfect CSI because of noise on channel estimation (\ref{ce_w_corr}), we denote with $\hat{R}_k(t)$ the estimate at BSs of the spectral efficiency achieved by UE $k$. We consider a constraint on the cluster size by assuming that each UE can be served by up to $J_{\rm MAX}$ BSs. Let us define the step function
\begin{equation}
\mathds{1}(x) =
\begin{cases}
1,& x > 0\,,\\
0,& x \leq 0\,.
\end{cases}
\end{equation}
Therefore, the weighted sum rate maximization can be written as
\begin{subequations}
\begin{equation}
\max_{ \mathcal{S}(t)\subseteq\mathcal{K} , \left\{\bm{G}_{k,j}(t)\right\}_{k\in\mathcal{S}(t),j\in\mathcal{J}} , \left\{\bm{P}_k(t)\right\}_{k\in\mathcal{S}(t)} } \sum_{k\in\mathcal{S}(t)} \alpha_k(t) \hat{R}_k(t)
\label{fo_g}
\end{equation}
s.t.
\begin{equation}
\sum_{k\in\mathcal{S}(t)} {\rm tr} \left( \bm{G}_{k,j}^{H}(t) \bm{G}_{k,j}(t) {\rm diag}\left(\bm{P}_k(t)\right) \right) \leq P^{({\rm BS})}\,,\quad j\in\mathcal{J}\,,
\label{power_constraints_g}
\end{equation}
\begin{equation}
\sum_{j\in\mathcal{J}} \mathds{1} \left( \left\| \bm{G}_{k,j}(t) \right\|^2 \right) \leq J_{\rm MAX} \,,\quad k\in\mathcal{S}(t)\,,
\label{cluster_constraints_g}
\end{equation}
\label{problem_g}
\end{subequations}
where scaling factor $\alpha_k(t)$ in (\ref{fo_g}) is the quality of service (QoS) for UE $k$ which depends on the employed scheduler, (\ref{power_constraints_g}) is the power constraint at each BS and (\ref{cluster_constraints_g}) imposes that each UE cannot be served by more than $J_{\rm MAX}$ BSs. Note that optimization (\ref{problem_g}) is very general and includes a) BS clustering, b) UE scheduling, c) beamforming design, d) transmission rank selection and e) power allocation. A practical solution to solve (\ref{problem_g}) is described in Section \ref{algorithm_section}.

\subsection{Third Phase: Downlink Data Transmission}

After computing the solution to problem (\ref{problem_g}), BS clusters serve the scheduled UEs by using the $N_E - N_T$ resource elements still available in block $t$. For the sake of clarity, in the rest of the paper we drop the block index $t$. By defining matrix $\bm{G}_k = \left[ \bm{G}_{k,1}^T , \bm{G}_{k,2}^T , \ldots , \bm{G}_{k,J}^T \right]^T$, with $\left\| \left[ \bm{G}_k \right]_{\cdot,l} \right\|^2=1$, $l=0,1,\ldots,l_k-1$, and matrix $\bm{H}_k = \left[ \bm{H}_{k,1} , \bm{H}_{k,2} , \ldots , \bm{H}_{k,J} \right]$, the signal received by UE $k$ can be written as
\begin{equation}
\bm{r}_k = \bm{H}_k \bm{G}_k \bm{s}_k + \sum_{m \in \mathcal{S} \setminus \{k\}} \bm{H}_k \bm{G}_m \bm{s}_m + \bm{n}_k\,,
\label{received signal}
\end{equation}
where $\bm{s}_k \sim \mathcal{CN}\left(\bm{0}_{l_k \times 1},{\rm diag}\left(\bm{P}_k\right)\right)$ is the data symbol vector transmitted toward UE $k$ and $\bm{n}_k \sim \mathcal{CN}\left(\bm{0}_{ N \times 1 } , \sigma_n^2 \bm{I}_N \right)$ is the thermal noise at the UE antennas. We assume perfect CSI at the UE side, which employs IRC with successive interference cancellation (SIC) \cite[Ch.~10]{tse_viswanath}. We assume perfect detection, hence there is no error propagation. Note that IRC both minimizes the mean square error and maximizes the SINR at the detection point \cite{irc_combiner_winters}. 

From (\ref{received signal}), by defining the interference plus noise covariance matrix at the UE as
\begin{equation}
\bm{\Psi}_k = \sigma_n^2 \bm{I}_N + \sum_{m \in \mathcal{S} \setminus \{k\}} \bm{H}_k \bm{G}_m {\rm diag}\left(\bm{P}_m\right) \bm{G}_m^H \bm{H}_k^H\,,
\label{psi_k}
\end{equation}
the effective spectral efficiency achieved by UE $k$ can be written as
\begin{equation}
\begin{split}
R_k = & \left( 1 - \frac{N_T}{N_E}\right) \times \\
& \log_2 {\rm det} \left( \bm{I}_N + \bm{H}_k \bm{G}_k {\rm diag}\left(\bm{P}_k\right) \bm{G}_k^H \bm{H}_k^H \bm{\Psi}_k^{-1} \right)\,,
\end{split}
\label{R_k}
\end{equation}
where in (\ref{R_k}) the overhead due to the UE pilot transmission is taken into account in the scaling factor before the logarithm.

\paragraph*{Remark} Note that the assumption of perfect CSI at the UE in the considered system does not limit the value of the developed analysis. Indeed, it has already been shown \cite{marzetta_apr06, gomodam_icc08} that in a similar setup the overhead required to obtain a reliable CSI at the UE is almost negligible when compared to the overhead necessary to acquire CSI at the BSs. This result is simply explained by the huge difference in terms of available power at UEs and BSs, which is 23 dB in the LTE scenario considered in Section \ref{numerical_section}. However, note that in our model UE $k$, in order to implement IRC, needs to know all the equivalent channels $\bm{H}_k \bm{G}_m$, $m\in\mathcal{S}$, which can be estimated by using orthogonal training sequences among the different BS clusters. Simulation results not reported here confirm the results already obtained in literature by showing that perfect CSI performance can be approached with a small number of downlink training symbols.

\section{Dynamic Clustering and Resource Allocation Algorithm}
\label{algorithm_section}

Problem (\ref{problem_g}) poses several issues. First, solving the problem at the CU means that the backhaul infrastructure should be able to provide the CSI from all the BSs to the CU, which can be problematic, in particular when the number $J$ of BSs is very high and some of them may be far away from the CU. In such a case a BS may be connected to the CU through a multi-hop link and latency may become non negligible. However, a detailed analysis of the impact of latency on system performance strongly depends on the backhaul network infrastructure and is beyond the scope of this work. Moreover, even by assuming that a reliable CSI can be collected by the CU, there is a complexity issue when the number of BSs and UEs is high.

Hence, in this work, we propose a practical solution to (\ref{problem_g}) where the computational burden can be partially distributed among the BSs. Let us define the function $f_k:\mathcal{J} \rightarrow \mathcal{J}$, which orders the BSs on the basis of the large scale fading component of the channel with respect to UE $k$, i.e., $\sigma_{k,f_k(c_1)}^2 > \sigma_{k,f_k(c_2)}^2$ if $c_1 < c_2$. Then, we indicate with $\mathcal{J}_k^{(u)}$ the cluster of $u$ BSs with the strongest average channel toward UE $k$, i.e.,
\begin{equation}
\mathcal{J}_k^{(u)} = \left\{ f_k(1) , f_k(2) , \ldots , f_k(u) \right\}\,.
\label{J_k_u}
\end{equation}
Hence, $f_k(1)$ is the anchor BS for UE $k$, i.e., the BS characterized by the highest average signal to noise ratio (SNR).

In a network with $J$ BSs and a maximum cluster size of $J_{\rm MAX}$, the number of possible BS clusters that can be constructed to serve a given UE is
\begin{equation}
\sum_{j=1}^{J_{\rm MAX}} \binom{J}{j}\,,
\label{c_clusters_number}
\end{equation}
which rapidly increases with $J$. However, as most of the interference at each UE comes from the closest BSs, we limit the number of candidate BS clusters that can be organized by the CU.
We order the BS clusters (\ref{J_k_u}) by an integer index $c\in\mathcal{C}$ and denote by $\mathcal{J}_c$ cluster $c$. By imposing a constraint on the maximum cluster size, we assume that $\mathcal{C}$ includes {\em all and only} the sets $\mathcal{J}_k^{(u)}$ whose size is not bigger than $J_{\rm MAX}$. Note that the considered assumption yields an important saving in terms of computational complexity by strongly limiting the number of candidate clusters with respect to (\ref{c_clusters_number}): this complexity saving is evaluated in Section \ref{numerical_section} for a typical LTE scenario.

For each cluster $\mathcal{J}_c$, we define the corresponding set $\mathcal{U}_c$ of UEs that can be scheduled for reception, which is formed by the UEs whose anchor BS belongs to $\mathcal{J}_c$, i.e.,
\begin{equation}
\mathcal{U}_c = \left\{ k \in \mathcal{K} : f_k(1) \in \mathcal{J}_c \right\}\,.
\label{U_c}
\end{equation}
Note that (\ref{U_c}) allows BSs in cluster $\mathcal{J}_c$ to serve all the UEs in its coverage area, even UEs close to the border. Although a different choice could be taken for instance by forcing the cluster to serve only the UEs far away from the border, it has been shown in \cite{baracca_iswcs12} that this alternative choice provides worse performance than (\ref{U_c}) when a huge network is considered and fairness among the UEs is taken into account.

Then, a solution to problem (\ref{problem_g}) is obtained by applying the following two-step algorithm.
\begin{enumerate}
	\item For each candidate BS cluster $\mathcal{J}_c$, we estimate the weighted sum rate $\hat{R}^{(c)}$ by selecting a suitable subset of UEs $\mathcal{S}_{c} \subseteq \mathcal{U}_c$, designing precoders, selecting transmission ranks and allocating powers. With the aim of allowing a distributed implementation of the algorithm, we perform this optimization without requiring any information from the other candidate BS clusters.
	\item After collecting the weighted sum rate $\hat{R}^{(c)}$ from all the candidate BS clusters in $\mathcal{C}$, the CU schedules a set of {\em non-overlapping} BS clusters, where each BS belongs to at most one cluster.
\end{enumerate}

First, although a solution with overlapping clusters would provide higher rates, it would be much more challenging in terms of computational complexity, because of the higher number of available cluster combinations. Moreover, it would require that everything is optimized at the CU \cite{gong_globecom11}. Hence, we focus here on the easier and more practical non-overlapping cluster solution.

Second, note that the proposed method can also be implemented in a fully centralized system, for instance by a star backhaul network topology \cite{boccardi_asilomar08}, where the CU is directly connected by a low latency link to each BS. However, this situation is unlikely when the number of BSs $J$ is high, and a more realistic scenario considers a backhaul network with direct links only among neighboring BSs \cite{biermann_wowmom11}. In such a case, the developed scheme properly adapts to the backhaul infrastructure and, for example, set $\mathcal{C}$ could be partitioned into subsets, each with a BS responsible for estimating the weighted sum rate achieved by all the candidate BS clusters of that subset. Then, this BS would forward only the estimated weighted sum rates to the CU which is managing the whole network.

Moreover, based on (\ref{U_c}) we observe that UE $k$ can be selected only by clusters that include its anchor BS $f_k(1)$. Hence, if we enforce a non-overlapping solution, each UE is never scheduled by two different non-overlapping clusters in the same block. However, we highlight that the proposed dynamic solution allows the flexibility of scheduling a given UE in different clusters across successive blocks.

In the rest of this section we describe more in detail the above two main steps of the algorithm. We stress that the candidate cluster selection, i.e., the construction of set $\mathcal{C}$ depends on the large scale fading: hence, in our model, it should be performed only every $T$ blocks. On the other hand, the two-step algorithm proposed to solve (\ref{problem_g}) follows a fast fading time-scale and therefore must be implemented in each block.

\subsection{Cluster Weighted Sum Rate Estimation}
\label{cluster_rate_section}

By defining the bijective function $\Upsilon_c:\left\{1,2,\ldots,\left|\mathcal{J}_c\right|\right\} \rightarrow \mathcal{J}$ which maps BSs in cluster $\mathcal{J}_c$ to set $\mathcal{J}$, we denote with a) $\hat{\bm{H}}_k^{(c)} = \left[ \hat{\bm{H}}_{k,\Upsilon_c(1)} , \hat{\bm{H}}_{k,\Upsilon_c(2)} , \ldots , \hat{\bm{H}}_{k,\Upsilon_c(\left|\mathcal{J}_c\right|)} \right]$ the matrix of size $N \times M\left|\mathcal{J}_c\right|$ collecting the MIMO channel estimated by BSs in $\mathcal{J}_c$ toward UE $k$ and b) $\bm{G}_k^{(c)} = \left[ \bm{G}_{k,\Upsilon_c(1)}^T , \bm{G}_{k,\Upsilon_c(2)}^T , \ldots , \bm{G}_{k,\Upsilon_c(\left|\mathcal{J}_c\right|)}^T \right]^T$ the precoding matrix of size $M\left|\mathcal{J}_c\right| \times l_k$ used by BSs in $\mathcal{J}_c$ to serve UE $k$. Due to the imperfect CSI at BSs and using (\ref{received signal}), the signal received by UE $k$ is modeled at cluster $\mathcal{J}_c$ as
\begin{equation}
\hat{\bm{r}}_k^{(c)} = \hat{\bm{H}}_k^{(c)} \bm{G}_k^{(c)} \bm{s}_k + \sum_{m \in \mathcal{S}_c \setminus \{k\}} \hat{\bm{H}}_k^{(c)} \bm{G}_m^{(c)} \bm{s}_m + \bm{n}_k + \hat{\bm{i}}_k^{(c)} \,,
\label{received signal_approx}
\end{equation}
where $\hat{\bm{i}}_k^{(c)}$ is the estimate of the inter-cluster interference (ICI) suffered by UE $k$. Note that the exact value of $\hat{\bm{i}}_k^{(c)}$ depends on the beamformers used by other clusters to serve their own UEs. Hence, an evaluation of $\hat{\bm{i}}_k^{(c)}$ would require an additional coordination among the clusters which should a) exchange CSI and b) jointly optimize UE selection, precoders and powers. However, with the aim of reducing the CSI exchange on the backhaul, we simply assume $\hat{\bm{i}}_k^{(c)} \sim \mathcal{CN}\left(\bm{0}_{ N \times 1 } , \xi_k^{(c)}\bm{I}_N \right)$ with
\begin{equation}
\xi_k^{(c)} = P^{({\rm BS})} \sum_{j \in \mathcal{J} \setminus \mathcal{J}_c} \sigma_{k,j}^2 \,.
\label{ici_power}
\end{equation}
Note that (\ref{ici_power}) represents the average ICI power at the UE $k$ when all the BSs outside cluster $c$ are transmitting at full power \cite[(2)]{huh_ici}.

Similarly to (\ref{psi_k}), from (\ref{received signal_approx}) we define the interference plus noise covariance matrix
\begin{equation}
\begin{split}
\hat{\bm{\Psi}}_k^{(c)} = & \left( \sigma_n^2 + \xi_k^{(c)} \right) \bm{I}_N + \\
& \sum_{m \in \mathcal{S}_c \setminus \{k\}} \hat{\bm{H}}_k^{(c)} \bm{G}_m^{(c)} {\rm diag}\left(\bm{P}_m\right) \bm{G}_m^{(c)H} \hat{\bm{H}}_k^{(c)H}\,.
\end{split}
\label{psi_k_approx}
\end{equation}
Last, BSs in $\mathcal{J}_c$ yield an estimate of the weighted sum rate $\hat{R}^{(c)}$ by solving the optimization problem (\ref{problem_g_cluster}), at the top of the page.
\begin{figure*}[!t]
\normalsize
\begin{subequations}
\begin{equation}
\begin{split}
\hat{R}^{(c)} = \max_{ \mathcal{S}_c\subseteq\mathcal{U}_c , \left\{\bm{G}_{k}^{(c)}\right\}_{k\in\mathcal{S}_c} , \left\{\bm{P}_k\right\}_{k\in\mathcal{S}_c} } & \sum_{k\in\mathcal{S}_c} \alpha_k \left( 1 - \frac{N_T}{N_E}\right) \times \\
& \log_2 {\rm det} \left( \bm{I}_N + \hat{\bm{H}}_k^{(c)} \bm{G}_k^{(c)} {\rm diag}\left(\bm{P}_k\right) \bm{G}_k^{(c)H} \hat{\bm{H}}_k^{(c)H} \hat{\bm{\Psi}}_k^{(c)-1} \right)
\end{split}
\label{fo_g_cluster}
\end{equation}
s.t.
\begin{equation}
\sum_{k\in\mathcal{S}_c} {\rm tr} \left( \bm{G}_{k,j}^{H} \bm{G}_{k,j} {\rm diag}\left(\bm{P}_k\right) \right) \leq P^{({\rm BS})} \,,\quad j\in\mathcal{J}_c\,.
\label{power_constraints_g_cluster}
\end{equation}
\label{problem_g_cluster}
\end{subequations}
\hrulefill
\vspace*{4pt}
\end{figure*}

Maximization (\ref{problem_g_cluster}) is a well studied multi-user MIMO problem \cite{mu_mimo} involving a) UE selection, b) transmission rank selection, c) precoding design and d) power allocation. Note that the estimate of the rate achieved by UE $k$ in (\ref{fo_g_cluster}) is computed by assuming that  the multiple receive antennas are exploited by performing SIC with IRC: the rank $l_k$ allocated to UE $k$ is given by the number of columns of precoder $\bm{G}_k^{(c)}$, whereas the remaining $N-l_k$ degrees of freedom are used to partially suppress the residual ICI. We propose to solve (\ref{problem_g_cluster}) under the following assumptions.

\begin{itemize}

\item Equal power is allocated to the streams sent toward the scheduled UEs, i.e., $P_{k,l} = P^{(c)}$, $k \in \mathcal{S}_c$, $l=0,1,\ldots,l_k-1$, where $P^{(c)}$ can be analytically computed from (\ref{power_constraints_g_cluster}) as
\begin{equation}
\displaystyle P^{(c)} = \frac{ P^{({\rm BS})}  }{\displaystyle \max_{j\in\mathcal{J}_c,} \sum_{k\in\mathcal{S}_c} \sum_{l=0}^{l_k-1} \left\| \left[ \bm{G}_{k,j} \right]_{\cdot,l} \right\|^2}\,.
\end{equation}

\item Beamformers are designed by using the multiuser eigenmode transmission (MET) scheme \cite{boccardi_icassp07}, where the precoding matrix used to serve UE $k$ is optimized with the aim of nullifying the interference toward the eigenmodes selected for the co-scheduled UEs $m\in\mathcal{S}_c\setminus\left\{k\right\}$. In detail, let $\hat{\bm{H}}_k^{(c)} = \hat{\bm{U}}_k^{(c)} \hat{\bm{\Sigma}}_k^{(c)} \hat{\bm{V}}_k^{(c)H}$ be the singular value decomposition (SVD) of matrix $\hat{\bm{H}}_k^{(c)}$, where the eigenvalues in $\hat{\bm{\Sigma}}_k^{(c)}$ are arranged so that the ones selected for transmission toward UE $k$ appear in the leftmost columns. By defining matrix
\begin{equation}
\begin{split}
\hat{\bm{\Gamma}}_k^{(c)} = & \left[ \left[ \hat{\bm{\Sigma}}_k^{(c)} \right]_{0,0} \left[ \hat{\bm{V}}_k^{(c)} \right]_{\cdot,0} , \left[ \hat{\bm{\Sigma}}_k^{(c)} \right]_{1,1} \left[ \hat{\bm{V}}_k^{(c)} \right]_{\cdot,1} , \ldots , \right.\\
& \left. \left[ \hat{\bm{\Sigma}}_k^{(c)} \right]_{l_k-1,l_k-1} \left[ \hat{\bm{V}}_k^{(c)} \right]_{\cdot,l_k-1} \right]^H\,,
\end{split}
\end{equation}
precoding matrix used to serve UE $k$ satisfies constraints
\begin{equation}
\hat{\bm{\Gamma}}_m^{(c)} \hat{\bm{G}}_k^{(c)} = \bm{0}_{l_m \times l_k} \,,\quad k \neq m \,.
\end{equation}

\item The eigenmodes (and accordingly the set $\mathcal{S}_c$ of scheduled UEs and the transmission rank allocated to each UE $k\in\mathcal{S}_c$) are selected by using a greedy iterative algorithm which, at each iteration, includes the eigenmode which maximizes the weighted sum rate $\hat{R}^{(c)}$  among the ones not scheduled in the previous iterations. The algorithm starts with no UE scheduled and stops when no increase in the weighted sum rate $\hat{R}^{(c)}$ is observed. Cluster $\mathcal{J}_c$, among the $N\left| \mathcal{U}_c \right|$ possible eigenmodes, selects a maximum of $M\left| \mathcal{J}_c \right|$ eigenmodes, due to the limited number of BS antennas. Note that the considered method flexibly adapts to the channel conditions by allowing the allocation of a) different ranks to different UEs in the same block and b) different ranks to the same UE across successive blocks.

\end{itemize}

\subsection{Clustering Optimization}
\label{clustering_section}

After collecting all the weighted sum rates $\hat{R}^{(c)}$, $c\in\mathcal{C}$, the CU schedules a set of non-overlapping clusters by maximizing the system weighted sum rate. In detail, by defining
\begin{equation*}
\begin{array}{rcl}
a_{j,c} & = & \begin{cases}
1,& \hbox{$j\in\mathcal{J}_c$,}\\
0,& \hbox{otherwise,}
\end{cases} \vspace{0.3cm} \\
x_{c} & = & \begin{cases}
1,& \hbox{CU schedules candidate cluster $\mathcal{J}_c$,}\\
0,& \hbox{otherwise,}
\end{cases}\\
\end{array}
\end{equation*}
we consider that each BS belongs to at most one cluster, i.e., we impose
\begin{equation}
\sum_{c\in\mathcal{C}} a_{j,c} \, x_c \leq 1\,,\quad j \in \mathcal{J} \,.
\label{nonoverl_cl}
\end{equation}
Therefore, at the CU the clustering optimization is performed by solving the following linear integer optimization problem
\begin{equation}
\max_{x_c ,\, c \in \mathcal{C}} \, \sum_{c\in\mathcal{C}} \, \hat{R}^{(c)} \, x_c\,,
\label{clustering_problem}
\end{equation}
\begin{center}
s.t. (\ref{nonoverl_cl}).
\end{center}

Note that (\ref{clustering_problem}) differs from the optimization carried out in \cite{weber_vtcspring11} where the objective function simply depends on the received power measured by the UEs.

Maximization (\ref{clustering_problem}) is the optimization version of the {\em set packing} problem, which is shown to be NP-hard \cite{set_packing_ref}. Hence, as the exhaustive search is not a viable method to solve (\ref{clustering_problem}), we propose a greedy iterative algorithm which is reported in Tab. \ref{alg_ref}: the proposed solution basically selects at each iteration the best (in terms of system weighted sum rate) cluster and ends when each BS has been assigned to at least one cluster. In detail, let $\mathcal{C}^{\rm(A)}(n)$ be the set of candidate clusters considered at iteration $n$. The algorithm starts by imposing $\mathcal{C}^{\rm(A)}(1) \gets \mathcal{C}$ and ends when $\mathcal{C}^{\rm(A)}(n)=\emptyset$. Note that $\mathcal{C}^{\rm(A)}(n)$ includes all the candidate clusters that do not overlap with the clusters scheduled in the previous iterations. At iteration $n$, we select cluster $\mathcal{J}_w \in \mathcal{C}^{\rm(A)}(n)$ that maximizes the per-BS weighted sum rate, i.e.,
\begin{equation}
\displaystyle \omega = \argmax_{c \in \mathcal{C}^{\rm(A)}(n)} \frac{\hat{R}^{(c)}}{\left| \mathcal{J}_c \right|} \,,
\label{J_w}
\end{equation}
and we remove from $\mathcal{C}^{\rm(A)}(n)$ all the clusters that partially overlap with $\mathcal{J}_w$. Note that in criterion (\ref{J_w}) we normalize the cluster weighted sum rate $\hat{R}^{(c)}$ with the number of BSs $\left| \mathcal{J}_c \right|$ included in the cluster with the aim of scheduling big clusters only if this really provides a system performance improvement. Let us consider as an example the basic scenario with $J=2$: by using (\ref{J_w}), we schedule the cluster of 2 BSs only if the weighted sum rate achieved within this cluster is higher than the system weighted sum rate achieved by the SCP scheme, i.e., when the 2 BSs are uncoordinated.

\begin{table}[t]
\centering
\caption{Greedy Iterative Algorithm for Cluster Selection (\ref{clustering_problem}).}
\label{alg_ref}
\hrulefill
\begin{algorithmic}[1]
\State $x_{c}^{(*)} \gets 0$, $c \in \mathcal{C}$
\State $n \gets 1$
\State $\mathcal{C}^{\rm(A)}(n) \gets \mathcal{C}$
\While{$\mathcal{C}^{\rm(A)}(n) \neq \emptyset$}
\State $\displaystyle \omega \gets \argmax_{c \in \mathcal{C}^{\rm(A)}(n)} \hat{R}^{(c)} / \left| \mathcal{J}_c \right|$
\State $x_{\omega}^{(*)} \gets 1$
\State $\displaystyle \mathcal{C}^{\rm(A)}(n+1) \gets \mathcal{C}^{\rm(A)}(n) \setminus \bigcup_{ \bar{c} \in \mathcal{C}^{\rm(A)}(n) : \mathcal{J}_{\bar{c}} \cap \mathcal{J}_{\omega} \neq \emptyset } \left\{ \bar{c} \right\} $
\State $n \gets n+1$
\EndWhile
\end{algorithmic}
\hrulefill
\end{table}

By denoting with $x_c^{(*)}$, $c\in\mathcal{C}$, the greedy solution to (\ref{clustering_problem}) obtained by applying the proposed algorithm, the set $\mathcal{S}$ of UEs scheduled in the current block turns out to be
\begin{equation}
\mathcal{S} = \bigcup_{c \in \mathcal{C} \,:\,x_c^{(*)}=1} \mathcal{S}_c\,.
\end{equation}
Finally, the precoding matrix and power used to serve UE $k\in\mathcal{S}$ are the ones computed within the cluster as described in Section \ref{cluster_rate_section}.

\section{Numerical Results}
\label{numerical_section}

\begin{figure}[t]
\centering
\includegraphics[width=0.7\hsize]{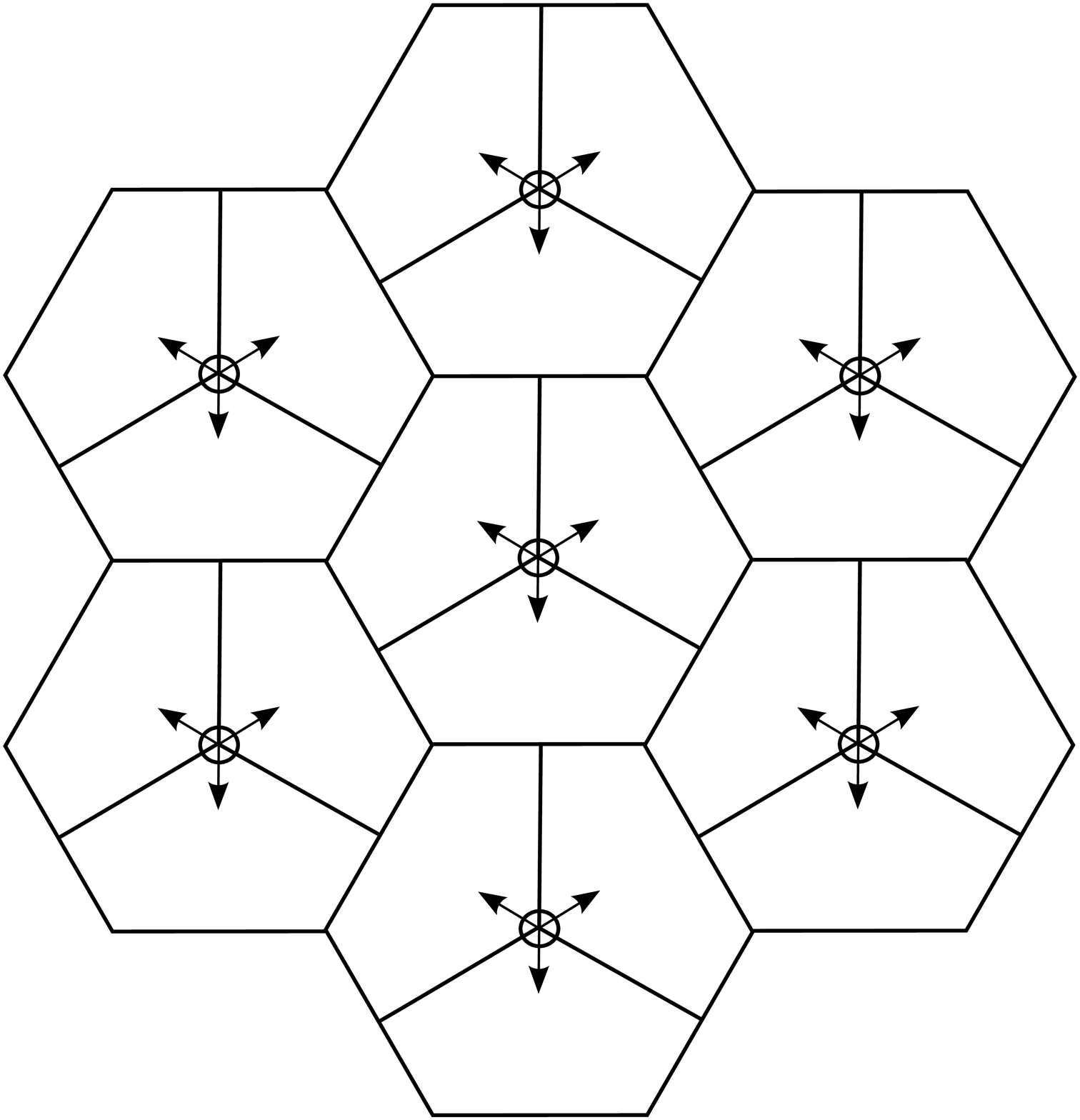}\\
\caption{Simulation setup with $J=21$ BSs organized in 7 sites, each with 3 sectors.}
\label{setup}
\end{figure}

We consider an hexagonal cellular scenario where $J=21$ BSs, each equipped with $M=4$ antennas, are organized in 7 sites, each with 3 co-located BSs (see also Fig. \ref{setup}). We consider 10 UEs randomly dropped in the coverage area of each BS, with $K=210$ UEs overall. The power available at each BS is $P^{({\rm BS})}=46$ dBm, the power available at each UE is $P^{({\rm UE})}=23$ dBm and the thermal noise power is $\sigma_n^2=-101$ dBm. The large scale fading between BS $j$ and UE $k$ can be written as
\begin{equation}
\sigma_{k,j}^2 = \Gamma^{({\rm CE})}  \left( \frac{d^{({\rm CE})}}{d_{k,j}} \right)^{\nu} \, e^{\zeta_{k,j}} \, A(\theta_{k,j})\,,
\end{equation}
where $d_{k,j}$ is the distance between BS $j$ and UE $k$, $\nu=3.5$ is the path-loss coefficient, $\Gamma^{({\rm CE})}\big|_{\rm dB} = 10$ dB is the average SNR when an UE is at the cell edge, $e^{\zeta_{k,j}}$ is the lognormal shadowing with 8 dB as standard deviation and $A(\theta_{k,j})$ models the antenna gain as a function of the direction $\theta_{k,j}$ of UE $k$ with respect to the antennas of BS $j$, with 
\begin{equation}
A(\theta_{k,j})\big|_{\rm dB} = -\min \left\{ 12 \left( \theta_{k,j} / \theta_{3{\rm dB}} \right)^2 ,  A_s \right\}\,,
\end{equation} 
where $\theta_{3dB} = (70/180)\pi$ and $A_s\big|_{\rm dB} = 20$ dB \cite[(21.3)]{lte_baker}. We consider an inter-site distance of 500 m and a minimum distance $d_{\rm min}=35$ m between BSs and UEs. Wraparound is used to deal with boundary effects \cite{wrap_ref}.
We also assume that channels are correlated by considering the popular Kronecker model \cite{mimo_kronecker_model}. By denoting with $\bm{R}_{\rm BS}$ the square correlation matrix of size $M$ at the BS, with ${\rm tr}(\bm{R}_{\rm BS})=M$, and with $\bm{R}_{\rm UE}$ the square correlation matrix of size $N$ at the UE, with ${\rm tr}(\bm{R}_{\rm UE})=N$, we can write
\begin{equation}
\bm{H}_{k,j}(t) = \bm{R}_{\rm UE}^{1/2} \bar{\bm{H}}_{k,j}(t) \left( \bm{R}_{\rm BS}^{1/2} \right)^{H}\,,
\label{kronecker_mimo_ch}
\end{equation}
where $\bar{\bm{H}}_{k,j}(t)$ is a matrix of size $N \times M$ whose entries are independent and identically distributed zero-mean complex Gaussian random variables with $\sigma_{k,j}^2$ as statistical power.

Results are obtained by simulating $100$ UE drops and $T=200$ block channel realizations for each UE drop. We assume that proportional fair scheduling \cite{tse_pfs} is implemented to provide fairness among UEs, i.e., $\alpha_k(t) = 1 / \bar{R}_k(t)$, with $\bar{R}_k(t+1)=(1-\gamma)\bar{R}_k(t) + \gamma R_k(t)$, $t=0,1,\ldots,T-1$, where $\gamma=0.1$ is the forgetting factor and we initialize $\bar{R}_k(1) = \log_2 \left( 1 + \bar{P}\sigma_{k,j_k}^2 / \sigma_n^2 \right)$. However, to allow the scheduler to reach a steady state, only the last $T/2$ channels of each UE drop are considered for system performance evaluation.

\begin{figure}[t]
\centering
\subfigure[c][ISC deployment.]{\includegraphics [width=0.4\hsize]{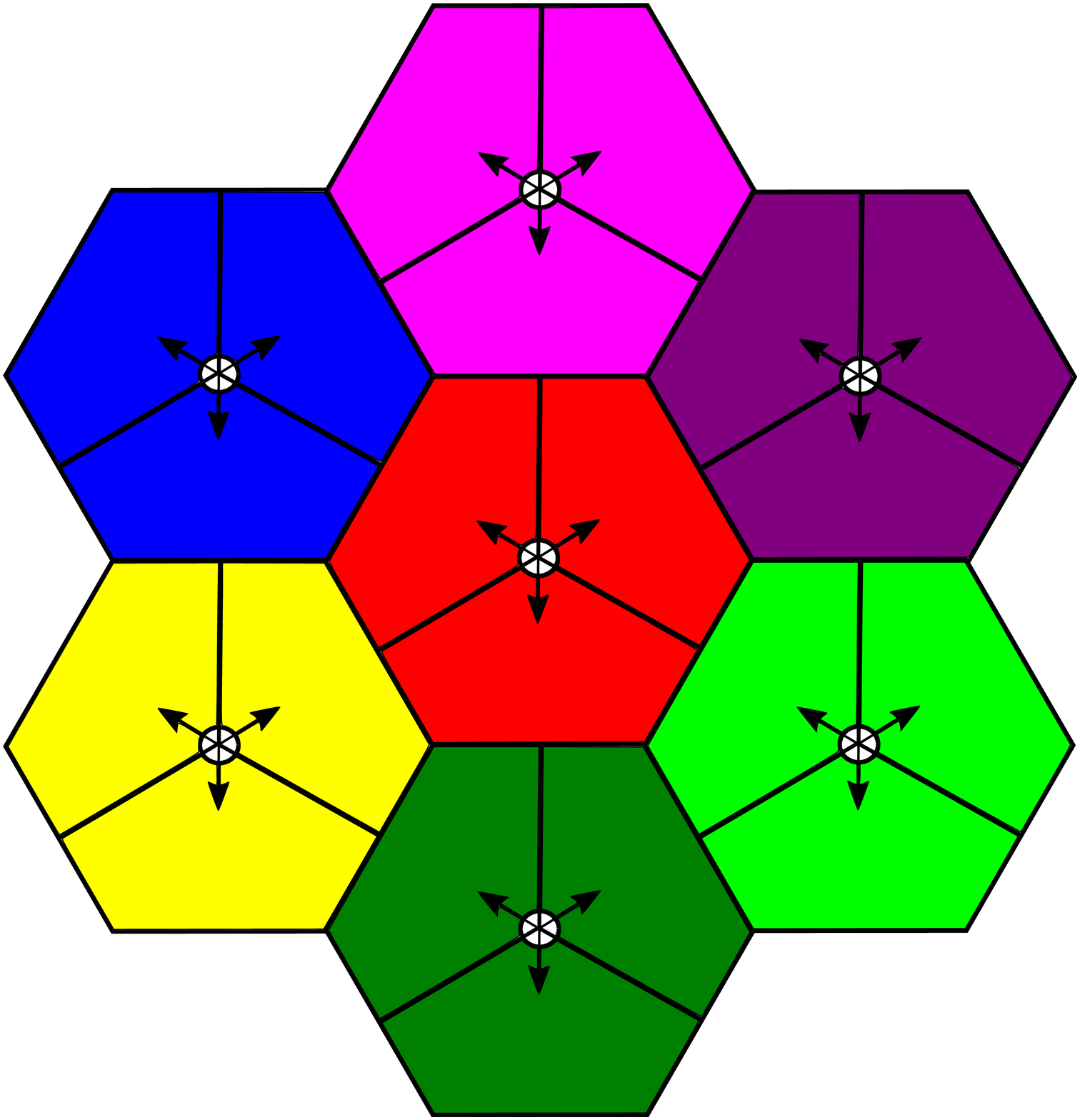}\label{isc_depl}}\qquad
\subfigure[c][SC deployment.]{\includegraphics [width=0.4\hsize]{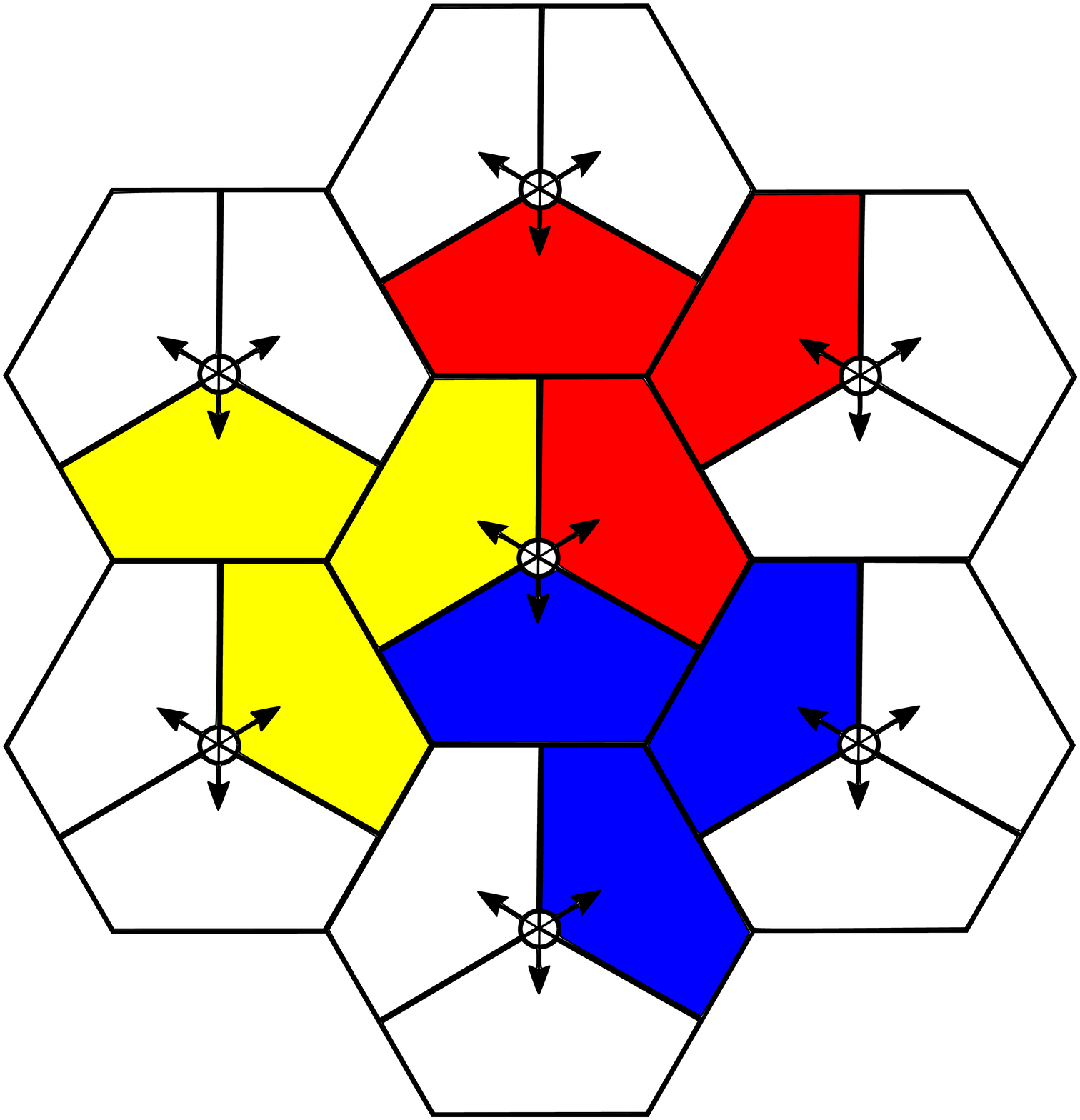}\label{sc_depl}}
\caption{Static CoMP-JP deployments considered in Section \ref{numerical_section}.}
\end{figure}

We compare the developed scheme based on dynamic clustering (DC) against:
\begin{itemize}
\item single cell processing (SCP), where no cooperation is allowed among the BSs and each UE is served by its anchor BS (baseline scheme);
\item intra-site cooperation (ISC), where we statically create 7 clusters, each one composed by 3 co-located BSs (see Fig. \ref{isc_depl}), and each UE is served by the best site in terms of average SNR;
\item static clustering (SC), where we still consider 7 static clusters, but with cooperation allowed among three BSs of three different sites as shown in Fig. \ref{sc_depl}.
\end{itemize}
Note that with ISC and SC there is no dynamic clustering, i.e., the clusters do not change over time adapting to the channel conditions. Moreover, we assume that UE scheduling, beamforming design, transmission rank selection and power allocation are performed by enforcing the assumptions described in Section \ref{cluster_rate_section} also with the static schemes: hence, UEs are served by using MET \cite{boccardi_icassp07} with equal power allocation among the eigenmodes and greedy user selection within each cluster.

The proposed schemes are compared in terms of:

\begin{itemize}

	\item UE rate, defined as
\begin{equation}
\bar{R}_k = \frac{2}{T} \sum_{t=T/2+1}^{T} R_k(t)\,,
\end{equation}

\item average cell rate, defined as
\begin{equation}
\bar{R}_{\rm cell} = \frac{1}{J} \sum_{k\in\mathcal{K}} \bar{R}_k\,.
\end{equation}

\end{itemize}

First, to evaluate the complexity saving achieved by the candidate cluster selection described at the beginning of Section \ref{algorithm_section}, we show in Tab. \ref{tab_n_cc} the 5th, the 50th, and the 95th percentiles of the number $\left|\mathcal{C}\right|$ of candidate clusters considered with DC by assuming $J_{\rm MAX}=3$. By adapting the candidate clusters to the long-term channel conditions with working assumption described at the beginning of Section \ref{algorithm_section}, we have a saving of about $80\%$ in terms of $\left|\mathcal{C}\right|$ with respect to the full search (\ref{c_clusters_number}): in fact, with our approach we ignore candidate clusters that include far apart BSs.

\begin{table}[t]
\centering
\caption{Number of candidate clusters with $J_{\rm MAX}=3$: comparison between DC and exhaustive search.}
\begin{tabular}{|c|c|c|c|c|}
\hline
 & DC-5th & DC-50th & DC-95th & eq. (\ref{c_clusters_number}) \\ \hline
 $\left|\mathcal{C}\right|$ & 235 & 249 & 263 & 1561 \\ \hline
\end{tabular}
\label{tab_n_cc}
\end{table}

\subsection{Effect of Multiple Antennas at the UEs}
\label{section_pCSI}

\begin{figure}[t]
\centering
\includegraphics[width=0.9\hsize]{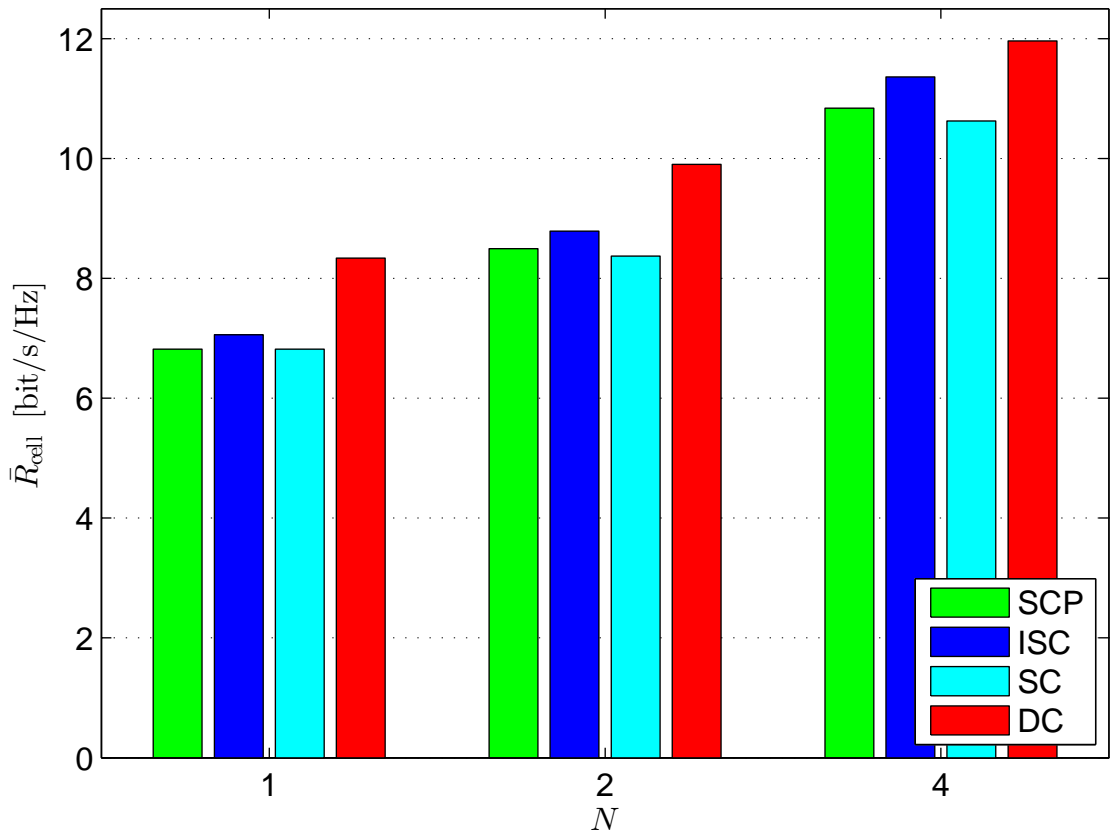}\\
\caption{Average cell rate for three values of $N$ and $M=4$.}
\label{fig_av_cell_rate_N}
\end{figure}

\begin{figure}[t]
\centering
\includegraphics[width=0.9\hsize]{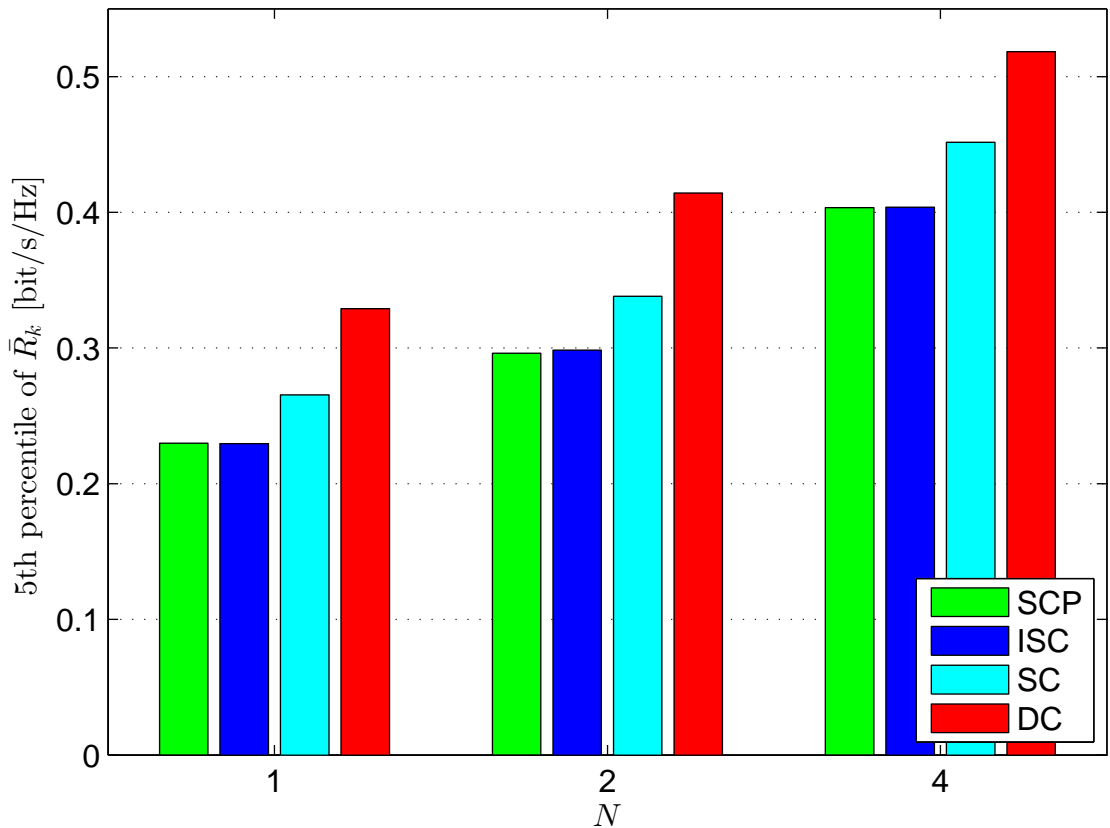}\\
\caption{5th percentile of the UE rate for three values of $N$ and $M=4$.}
\label{fig_fifth_ue_rate_N}
\end{figure}

In this section we consider perfect CSI at the BSs, i.e., $\hat{\bm{H}}_{k,j}(t)=\bm{H}_{k,j}(t)$ in (\ref{ce_w_corr}), uncorrelated antennas, i.e., in (\ref{kronecker_mimo_ch}) $\bm{R}_{\rm BS}=\bm{I}_M$ and $\bm{R}_{\rm UE}=\bm{I}_N$, and we assume $J_{\rm MAX}=3$ with DC for a fair comparison against ISC and SC in terms of maximum cluster size. In Fig.s \ref{fig_av_cell_rate_N} and \ref{fig_fifth_ue_rate_N} we report the average cell rate and the 5th percentile of the UE rate for three values of the number $N$ of UE antennas, respectively. First, we observe an important performance improvement by adding antennas at the UE side. 
For instance, with SCP by increasing $N$ from 1 to 4, there is a gain of about $76\%$ in terms of the 5th percentile of the UE rate.
Two factors mainly contribute to this gain: a) UEs with lower SINR use IRC to limit the impact of residual ICI not managed at the transmit side and b) UEs with higher SINR can be served by multiple streams of data.
From Tab. \ref{tab_ranks}, where we report the distribution of the number of streams $l_k$ with $N=4$, we note that with SCP more than $80\%$ of the transmissions are rank-1. On the other hand, with DC, as the interference level suffered by the UEs is lower, about $33\%$ of transmissions are multi-stream. This shows that in general most of the gain is due to the IRC and multi-stream transmission plays a non-negligible role only with DC. 
Then, we observe that ISC provides a moderate gain with respect to SCP in terms of cell rate, but almost no gain in terms of the 5th percentile of the UE rate, whereas the opposite happens with SC.
Indeed ISC of Fig. \ref{isc_depl}, by only allowing cooperation among the sectors of the same site, partially helps the UEs close to the site border, which however get better performance with SC of Fig. \ref{sc_depl}.
Moreover, we also observe that the performance gain achieved by DC over SCP decreases by adding more antennas at the UE side.
In fact, as the gain of using multiple antenna UEs is mainly due to the IRC which cancels ICI, the benefits of increasing $N$ are seen more in a non-cooperative scenario, where the residual ICI is higher with respect to DC.
In detail, from Fig. \ref{fig_fifth_ue_rate_N}, the performance gain achieved by DC over SCP drops from about $43\%$ with $N=1$ to about $28\%$ with $N=4$.

\begin{table}[t]
\centering
\caption{Distribution ($\%$) of $l_k$ with $N=4$.}
\begin{tabular}{|c||c|c|c|c|}
\hline
 & $l_k=1$ & $l_k=2$ & $l_k=3$ & $l_k=4$ \\ \hline\hline
 SCP & 82.6 & 16.0 & 1.4 & 0.0 \\ \hline
 ISC & 82.8 & 15.5 & 1.6 & 0.1 \\ \hline
 SC  & 82.9 & 15.7 & 1.4 & 0.0 \\ \hline
 DC  & 67.4 & 21.4 & 8.1 & 3.1 \\ \hline
\end{tabular}
\label{tab_ranks}
\end{table}

To further understand the role of IRC and multi-stream transmission with both SCP and CoMP, we consider now a simplified solution to (\ref{problem_g_cluster}) obtained by assuming a maximum number of streams $l^{({\rm MAX})}$ that can be transmitted toward each UE and limiting the eigenmodes that can be used for each UE to only the strongest $l^{({\rm MAX})}$ eigenmodes. In Tab. \ref{tab_rates_lMAX} we report the average cell rate and the 5th percentile of the UE rate by considering $N=4$ and $l^{({\rm MAX})}=1,4$. Note that $l^{({\rm MAX})}=1$ means that each UE is always served by rank-1 transmissions along its strongest eigenmode and implements IRC, whereas $l^{({\rm MAX})}=N=4$ means that there are no constraints on the number of data streams. Again, as the level of ICI is higher with SCP, in terms of the 5th percentile of the UE rate the gain achieved by DC over SCP decreases from about $28\%$ with $l^{({\rm MAX})}=4$ to about $18\%$ with $l^{({\rm MAX})}=1$. These results confirm that the gain of CoMP with respect to the baseline non-cooperative scheme can still be important when UEs are equipped with multiple antennas, but only if multi-stream transmission is properly exploited.

\begin{table}[t]
\centering
\caption{Average cell rate and 5th percentile of the UE rate with $N=4$ and $l^{({\rm MAX})}=1,4$.}
\begin{tabular}{|c||c|c||c|c|}
\hline
 & \multicolumn{2}{|c||}{$\bar{R}_{\rm cell}$ [bit/s/Hz]} & \multicolumn{2}{|c|}{5th percentile of $\bar{R}_k$ [bit/s/Hz]} \\ \hline
 & $l^{({\rm MAX})}=1$ & $l^{({\rm MAX})}=4$ & $l^{({\rm MAX})}=1$ & $l^{({\rm MAX})}=4$ \\ \hline\hline
 SCP & 10.74 & 10.84 & 0.415 & 0.404 \\ \hline
 DC  & 11.89 & 11.96 & 0.491 & 0.518 \\ \hline
\end{tabular}
\label{tab_rates_lMAX}
\end{table}

\subsection{Effect of Antenna Correlation}

\begin{figure}[t]
\centering
\includegraphics[width=0.9\hsize]{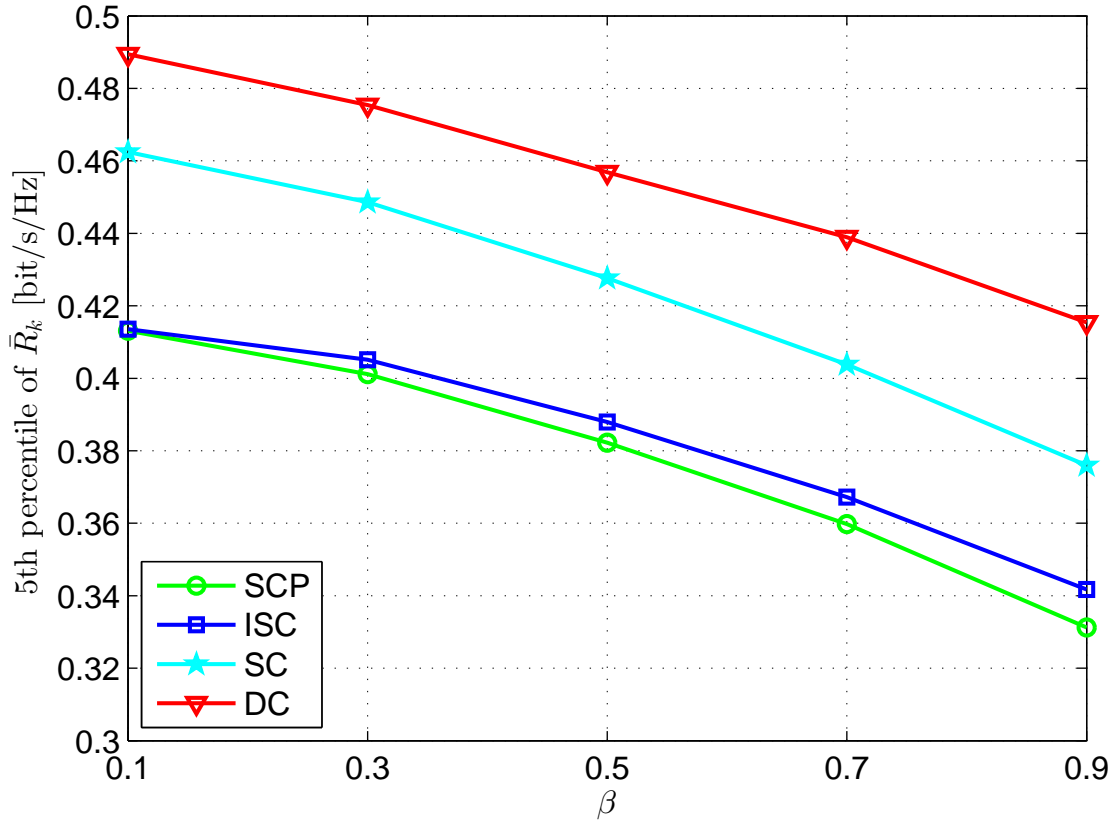}\\
\caption{5th percentile of the UE rate with respect to $\beta$ with $N=4$ and $l^{({\rm MAX})}=1$.}
\label{fig_beta}
\end{figure}

In Fig. \ref{fig_beta} we consider $N=4$, $l^{({\rm MAX})}=1$, and perfect CSI at BSs, and we introduce correlation among UE antennas by assuming that $\bm{R}_{\rm UE}$ is a symmetric Toeplitz matrix whose first column is $\left[\bm{R}_{\rm UE}\right]_{\cdot,0} = \left[1,\beta,\ldots, \beta^{N-1}\right]^T$, and plot the  5th percentile of the UE rate in terms of $\beta$.
As expected, a higher rate is achieved with low-correlated antennas, i.e., for lower values of $\beta$.
Moreover, we also observe that the gain achieved by DC over SCP decreases by decreasing $\beta$: in detail, this gain drops from $25\%$ with $\beta=0.9$ to $18\%$ with $\beta=0.1$.
In fact, by decreasing the correlation among UE antennas, we improve the interference suppression capability of IRC.
These results confirm that it is not worthy to add more antennas at the UE when they are strongly correlated.

\subsection{Effect of Cluster Size}

\begin{figure}[t]
\centering
\includegraphics[width=0.9\hsize]{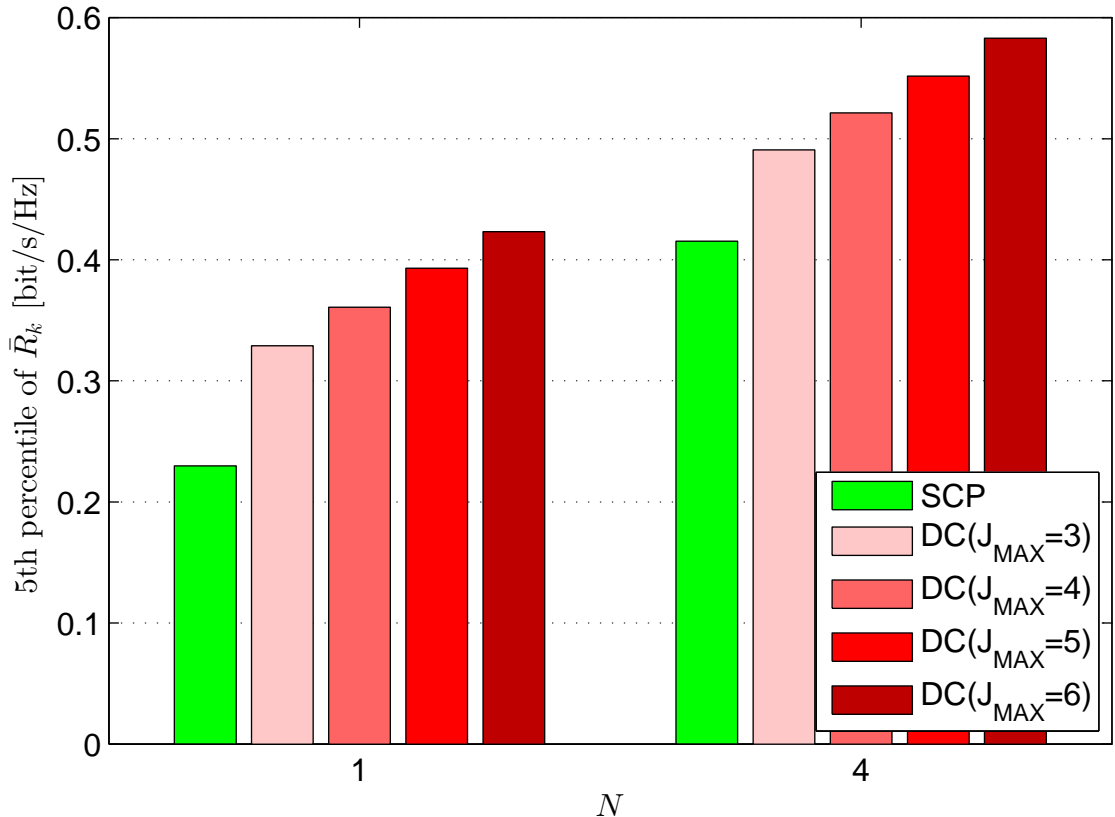}\\
\caption{5th percentile of the UE rate for two values of $N$ and four values of $J_{\rm MAX}$: $l^{({\rm MAX})}=1$.}
\label{fig_fifth_ue_rate_Jmax}
\end{figure}

In Fig. \ref{fig_fifth_ue_rate_Jmax}, by assuming $N=1,4$, $l^{({\rm MAX})}=1$, uncorrelated antennas and perfect CSI at BSs, we compare SCP and DC in terms of the 5th percentile of the UE rate for four values of the maximum cluster size $J_{\rm MAX}$. An important gain is observed with CoMP by increasing $J_{\rm MAX}$: for instance, the gain achieved by DC over SCP increases from $43\%$ ($18\%$) with $J_{\rm MAX}=3$ to about $84\%$ ($40\%$) with $J_{\rm MAX}=6$ when $N=1$ ($N=4$). These results show that although the strongest interferers are managed by CoMP with $J_{\rm MAX}=3$, the ICI suffered by UEs is still very high and strongly limits system performance. Hence, a general comment is that BS clusters of higher dimension should be employed if the backhaul infrastructure allows to do this.

\subsection{Effect of Imperfect CSI at the BSs}
\label{section_IpCSI}

In this section we assume that the CSI at the BSs is affected by noise (\ref{ce_w_corr}) due to the finite number of resource elements $N_T$ allocated to the pilot transmissions in each block.
After denoting with $f_d$ the maximum Doppler frequency and with $\bar{\tau}_{rms}$ the root-mean square delay spread of the channel, we define, respectively, the coherence bandwidth \cite[Ch.~4]{benv_02} and coherence time \cite[Ch.~4]{Rappaport} of the channel as 
\begin{subequations}
\begin{equation}
W_C = \frac{1}{\bar{\tau}_{rms}}\,,
\end{equation}
\begin{equation}
T_C = \frac{0.423}{f_d}\,.
\end{equation}
\end{subequations}
Note that above expressions are only used to determine the block size $N_E$ such that the channel can be modeled as uncorrelated between adjacent blocks. Indeed, if $f_d$ or $\bar{\tau}_{rms}$ increase, $N_E$ is reduced and this lowers the rate of each UE as given by (\ref{R_k}).
Due to the issues in obtaining a reliable CSI at the BSs in a high mobility scenario, in the following we consider $f_d = 5$ Hz, which roughly corresponds to a mobile velocity of 2 km/h \cite[Ch.~20]{lte_baker}.
In this section we also assume $N=4$, $l^{({\rm MAX})}=1$, uncorrelated antennas and $J_{\rm MAX}=3$ with DC.

\begin{figure}[t]
\centering
\includegraphics[width=0.9\hsize]{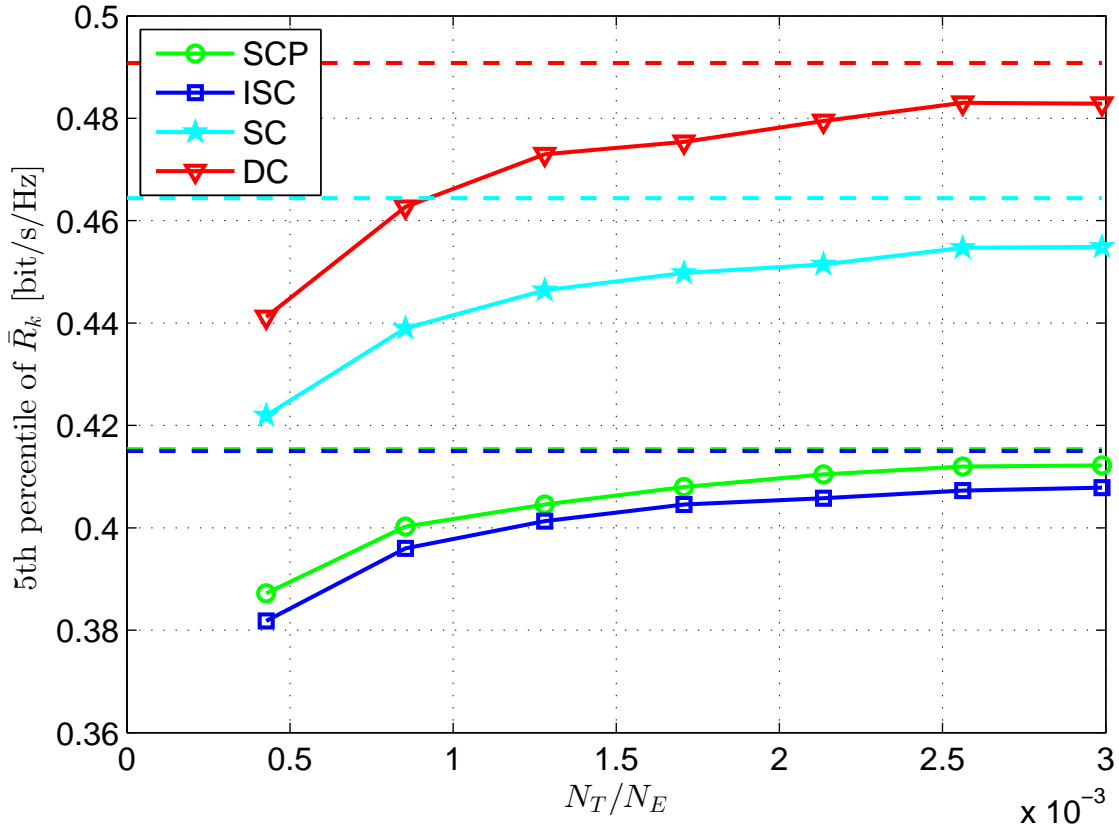}\\
\caption{5th percentile of the UE rate with respect to $N_T/N_E$ with $N=4$, $l^{({\rm MAX})}=1$, and for the EPA channel. The dashed lines are the rates computed in Section \ref{section_pCSI} by assuming perfect CSI at the BSs.}
\label{fig_epa}
\end{figure}

In Fig. \ref{fig_epa} we consider the extended pedestrian A (EPA) model, which is a very low frequency selective channel with $\bar{\tau}_{rms}=43$ ns \cite[Tab. 20.2]{lte_baker}. In detail, we show the 5th percentile of the UE rate in terms of the ratio $N_T/N_E$, which represents the fraction of resources used for pilot transmission. We observe that in this case, rate performance close to the perfect CSI case can be achieved by properly increasing the value of $N_T$. Moreover, note that SCP approaches the best performance faster than CoMP schemes. In fact, while with SCP only the channels between a BS and its anchored UEs are used for precoding design, with CoMP precoders are optimized on the basis also of the channels between some other auxiliary BSs and these UEs. As these channels are generally characterized by a lower SNR with respect to the channel between a BS and its anchored UEs, more pilots are necessary to collect a reliable CSI at transmit side.

\begin{figure}[t]
\centering
\includegraphics[width=0.9\hsize]{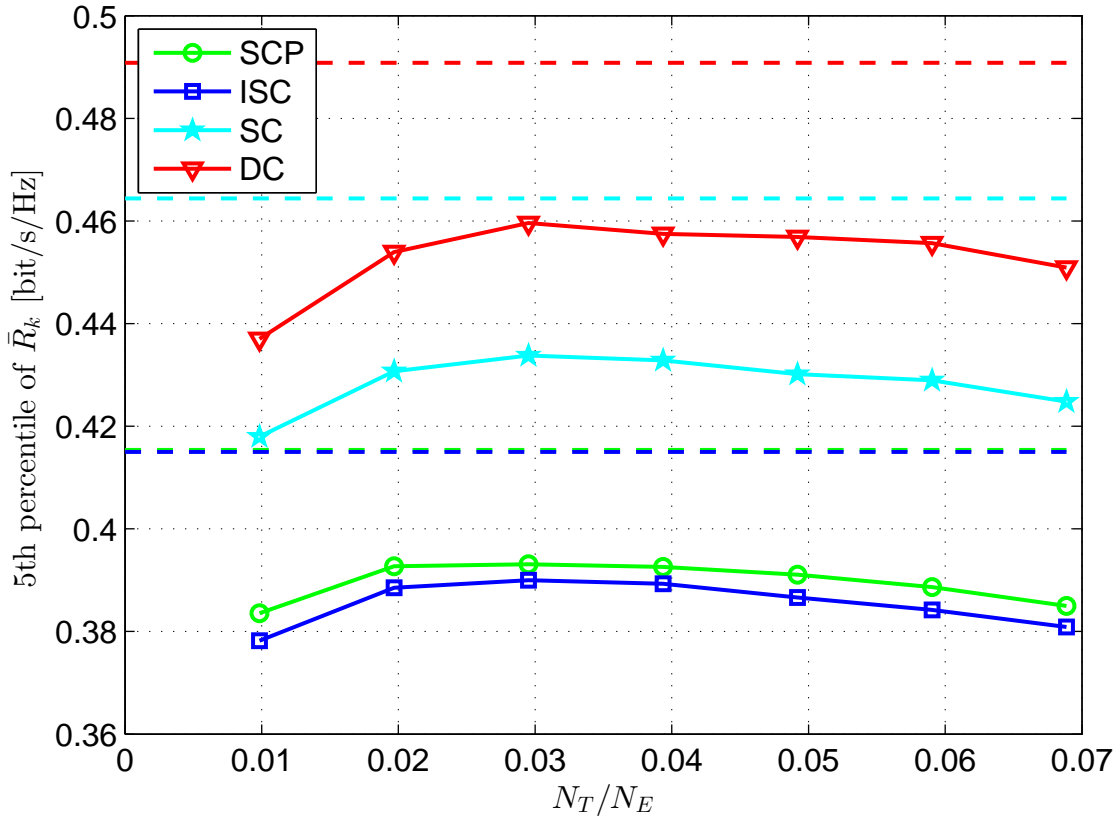}\\
\caption{5th percentile of the UE rate with respect to $N_T/N_E$ with $N=4$, $l^{({\rm MAX})}=1$, and for the ETU channel. The dashed lines are the rates computed in Section \ref{section_pCSI} by assuming perfect CSI at the BSs.}
\label{fig_etu}
\end{figure}

In Fig. \ref{fig_etu} we plot the 5th percentile of the UE rate with respect to the ratio $N_T/N_E$ for the very frequency selective extended typical urban (ETU) channel model, characterized by $\bar{\tau}_{rms}=991$ ns \cite[Tab. 20.2]{lte_baker}. In this case we observe that the rates increase with $N_T$ up to a maximum and then they start decreasing. In fact, increasing the value of $N_T$ has two conflicting effects: a) from (\ref{ce_wo_corr}) a more reliable CSI is collected at the BSs thus improving performance and b) a lower number of resource elements is allocated to data transmission thus obviously reducing the achievable rate. Clearly, for lower values of $N_T$ the effect of a better CSI dominates, whereas for higher values of $N_T$ the CSI is reliable enough for the SINR level of the UEs, and a further increase of the number of pilots represents only a waste of resources. Even if we are still considering a low mobility scenario, due to the higher frequency selectivity of the ETU channel model, no scheme reaches the rates achieved with perfect CSI. Then, as observed for the EPA model, the fraction of resources allocated to pilots necessary to reach the peaks is lower for SCP ($N_T/N_E \approx 0.02$) with respect to DC ($N_T/N_E \approx 0.03$). As a consequence, by choosing for each scheme the value of $N_T$ which provides best rate, the gain achieved by DC over SCP decreases with respect to the perfect CSI case to about $16\%$.

\section{Conclusions}
\label{conclusion_section}

In this paper we have considered a downlink CoMP-JP system and, by assuming a maximum cluster size, we have developed a dynamic clustering and resource allocation algorithm where the clusters change over time adapting to the channel conditions. We consider that UEs are equipped with multiple antennas that can implement IRC and be served by means of a multi-stream transmission. The proposed algorithm first defines a set of candidate BS clusters depending on the large scale fading. Then, a two step procedure is applied following a fast fading time scale: a) first, a weighted sum rate is estimated within each candidate cluster by performing UE selection and precoding, power and transmission rank selection, and then b) the CU schedules only a subset of these candidates by maximizing the system weighted sum rate. We highlight that joint optimization of transmitter and receiver is performed in the developed resource allocation scheme. Numerical results show that much higher rates can be achieved when UEs are equipped with multiple antennas. In fact, by reducing the level of interference suffered by UEs, the proposed approach exploits more the multi-stream transmission than SCP. However, as most of the gain is due to the IRC, the gain achieved by the proposed approach decreases with respect to SCP by increasing the number of UE antennas. Finally, when channel estimation is considered at the BSs, the gain promised in the perfect CSI scenario may be achieved only in part: in fact, a better estimate requires a longer training sequence and this lowers the system rate.

\section*{Acknowledgment}

Part of this work has been performed in the framework of the FP7 project ICT-317669 METIS, which is partly funded by the European Union. The authors would like to acknowledge the contributions of their colleagues in METIS, although the views expressed are those of the authors and do not necessarily represent the project.

\bibliographystyle{IEEEtran}
\bibliography{IEEEabrv,full_bibliography}

\begin{thebibliography}{10}
\providecommand{\url}[1]{#1}
\csname url@samestyle\endcsname
\providecommand{\newblock}{\relax}
\providecommand{\bibinfo}[2]{#2}
\providecommand{\BIBentrySTDinterwordspacing}{\spaceskip=0pt\relax}
\providecommand{\BIBentryALTinterwordstretchfactor}{4}
\providecommand{\BIBentryALTinterwordspacing}{\spaceskip=\fontdimen2\font plus
\BIBentryALTinterwordstretchfactor\fontdimen3\font minus
  \fontdimen4\font\relax}
\providecommand{\BIBforeignlanguage}[2]{{%
\expandafter\ifx\csname l@#1\endcsname\relax
\typeout{** WARNING: IEEEtran.bst: No hyphenation pattern has been}%
\typeout{** loaded for the language `#1'. Using the pattern for}%
\typeout{** the default language instead.}%
\else
\language=\csname l@#1\endcsname
\fi
#2}}
\providecommand{\BIBdecl}{\relax}
\BIBdecl

\bibitem{comp_book}
P.~Marsch and G.~Fettweis, \emph{Coordinated multi-point in mobile
  communications}.\hskip 1em plus 0.5em minus 0.4em\relax Cambridge University
  Press, 2011.

\bibitem{karakayali_aug06}
M.~K. Karakayali, G.~J. Foschini, and R.~A. Valenzuela, ``Network coordination
  for spectrally efficient communications in cellular systems,'' \emph{{IEEE}
  Wireless Commun. Mag.}, vol.~13, no.~4, pp. 56--61, Aug. 2006.

\bibitem{comp_tutorial}
D.~Gesbert, S.~Hanly, H.~Huang, S.~Shamai, O.~Simeone, and W.~Yu, ``Multi-cell
  {MIMO} cooperative networks: a new look at interference,'' \emph{{IEEE} J.
  Sel. Areas Commun.}, vol.~28, no.~9, pp. 1380--1408, Dec. 2010.

\bibitem{bjornson_book}
E.~Bj{\"{o}}rnson and E.~Jorswieck, ``Optimal resource allocation in
  coordinated multi-cell systems,'' \emph{Foundations and Trends in
  Communications and Information Theory}, vol.~9, no. 2-3, pp. 113--381, 2012.

\bibitem{3gpp_comp}
{3GPP TR 36.819 v11.1.0}, \emph{{Coordinated multi-point operation for LTE
  physical layer aspects (Release 11)}}, Dec. 2011.

\bibitem{hp_mayer}
R.~Irmer, H.~Droste, P.~Marsch, M.~Grieger, G.~Fettweis, S.~Brueck, H.-P.
  Mayer, L.~Thiele, and V.~Jungnickel, ``Coordinated multipoint: concepts,
  performance, and field trial results,'' \emph{{IEEE} Commun. Mag.}, vol.~49,
  no.~2, pp. 102--111, Feb. 2011.

\bibitem{zakhour_tsp11}
R.~Zakhour and D.~Gesbert, ``Optimized data sharing in multicell {MIMO} with
  finite backhaul capacity,'' \emph{{IEEE} Trans. Signal Process.}, vol.~59,
  no.~12, pp. 6102--6111, Dec. 2011.

\bibitem{baracca_const_quantization}
P.~Baracca, S.~Tomasin, and N.~Benvenuto, ``Constellation quantization in
  constrained backhaul downlink network {MIMO},'' \emph{{IEEE} Trans. Commun.},
  vol.~60, no.~3, pp. 830--839, Mar. 2012.

\bibitem{heath_tcom09}
J.~Zhang, R.~Chen, J.~G. Andrews, A.~Ghosh, and R.~W. Heath, ``Networked {MIMO}
  with clustered linear precoding,'' \emph{{IEEE} Trans. Wireless Commun.},
  vol.~8, no.~4, pp. 1910--1921, Apr. 2009.

\bibitem{papadogiannis_icc08}
A.~Papadogiannis, D.~Gesbert, and E.~Hardouin, ``A dynamic clustering approach
  in wireless networks with multi-cell cooperative processing,'' in \emph{Proc.
  IEEE International Conference on Communications (ICC)}, Beijing (China), May
  2008.

\bibitem{boccardi_asilomar08}
F.~Boccardi, H.~Huang, and A.~Alexiou, ``Network {MIMO} with reduced backhaul
  requirements by {MAC} coordination,'' in \emph{Proc. IEEE Conference on
  Signals, Systems and Computers (Asilomar)}, Pacific Grove (CA), Oct. 2008.

\bibitem{moon_vtc11}
J.-M. Moon and D.-H. Cho, ``Inter-cluster interference management based on
  cell-clustering in network {MIMO} systems,'' in \emph{Proc. IEEE Vehicular
  Technology Conference (VTC Spring)}, Budapest (Hungary), May 2011.

\bibitem{liu_wcsp09}
J.~Liu and D.~Wang, ``An improved dynamic clustering algorithm for multi-user
  distributed antenna system,'' in \emph{Proc. IEEE International Conference on
  Wireless Communications $\&$ Signal Processing (WCSP)}, Nanjing (China), Nov.
  2009.

\bibitem{zhou_globecom09}
S.~Zhou, J.~Gong, Z.~Niu, Y.~Jia, and P.~Yang, ``A decentralized framework for
  dynamic downlink base station cooperation,'' in \emph{Proc. IEEE Global
  Communications Conference (GLOBECOM)}, Honolulu (HI), Dec. 2009.

\bibitem{weber_vtcspring11}
R.~Weber, A.~Garavaglia, M.~Schulist, S.~Brueck, and A.~Dekorsy,
  ``Self-organizing adaptive clustering for cooperative multipoint
  transmission,'' in \emph{Proc. IEEE Vehicular Technology Conference (VTC
  Spring)}, Budapest (Hungary), May 2011.

\bibitem{papadogiannis_jan11}
A.~Papadogiannis, H.~J. Bang, D.~Gesbert, and E.~Hardouin, ``Efficient
  selective feedback design for multicell cooperative networks,'' \emph{{IEEE}
  Trans. Veh. Technol.}, vol.~60, no.~1, pp. 196--205, Jan. 2011.

\bibitem{gong_globecom11}
J.~Gong, S.~Zhou, Z.~Niu, L.~Geng, and M.~Zheng, ``Joint scheduling and dynamic
  clustering in downlink cellular networks,'' in \emph{Proc. IEEE Global
  Communications Conference (GLOBECOM)}, Houston (TX), Dec. 2011.

\bibitem{zakhour_aug10}
R.~Zakhour and D.~Gesbert, ``Distributed multicell-{MISO} precoding using the
  layered virtual {SINR} framework,'' \emph{{IEEE} Trans. Wireless Commun.},
  vol.~9, no.~8, pp. 2444--2448, Aug. 2010.

\bibitem{boccardi_2012}
F.~Boccardi, B.~Clerckx, A.~Ghosh, E.~Hardouin, G.~J{\"{o}}ngren, K.~Kusume,
  E.~Onggosanusi, and Y.~Tang, ``Multiple-antenna techniques in
  {LTE}-advanced,'' \emph{{IEEE} Commun. Mag.}, vol.~50, no.~3, pp. 114--121,
  Mar. 2012.

\bibitem{hwang-feb13}
I.~Hwang, C.-B. Chae, J.~Lee, and R.~W. Heath, ``Multicell cooperative systems
  with multiple receive antennas,'' \emph{{IEEE} Wireless Commun. Mag.},
  vol.~20, no.~1, pp. 50--58, Feb. 2013.

\bibitem{clercks_apr13}
B.~Clerckx, H.~Lee, Y.-J. Hong, and G.~Kim, ``A practical cooperative multicell
  {MIMO}-{OFDMA} network based on rank coordination,'' \emph{{IEEE} Trans.
  Wireless Commun.}, vol.~12, no.~4, pp. 1481--1491, Apr. 2013.

\bibitem{irc_combiner_winters}
J.~Winters, ``Optimum combining in digital mobile radio with cochannel
  interference,'' \emph{{IEEE} J. Sel. Areas Commun.}, vol.~2, no.~4, pp.
  528--539, Jul. 1984.

\bibitem{Kay-estimation}
S.~Kay, \emph{Fundamentals of statistical signal processing, volume I:
  estimation theory}.\hskip 1em plus 0.5em minus 0.4em\relax Prentice Hall,
  1993.

\bibitem{tse_viswanath}
D.~Tse and P.~Viswanath, \emph{Fundamentals of wireless communication}.\hskip
  1em plus 0.5em minus 0.4em\relax Cambridge University Press, 2005.

\bibitem{marzetta_apr06}
T.~Marzetta and B.~M. Hochwald, ``Fast transfer of channel state information in
  wireless systems,'' \emph{{IEEE} Trans. Signal Process.}, vol.~54, no.~4, pp.
  1268--1278, Apr. 2006.

\bibitem{gomodam_icc08}
K.~S. Gomadam, H.~C. Papadopoulos, and C.-E.~W. Sundberg, ``Techniques for
  multi-user {MIMO} with two-way training,'' in \emph{Proc. IEEE International
  Conference on Communications (ICC)}, Beijing (China), May 2008.

\bibitem{baracca_iswcs12}
P.~Baracca, F.~Boccardi, and V.~Braun, ``A dynamic joint clustering scheduling
  algorithm for downlink {CoMP} systems with limited {CSI},'' in \emph{Proc.
  IEEE International Symposium on Wireless Communication Systems (ISWCS)},
  Paris (France), Aug. 2012.

\bibitem{biermann_wowmom11}
T.~Biermann, L.~Scalia, C.~Choi, H.~Karl, and W.~Kellerer, ``Backhaul network
  pre-clustering in cooperative cellular mobile access networks,'' in
  \emph{Proc. IEEE International Symposium on a World of Wireless Mobile and
  Multimedia Networks (WoWMoM)}, Lucca (Italy), Jun. 2011.

\bibitem{huh_ici}
H.~Huh, A.~M. Tulino, and G.~Caire, ``Network {MIMO} with linear zero-forcing
  beamforming: large system analysis, impact of channel estimation, and
  reduced-complexity scheduling,'' \emph{{IEEE} Trans. Inf. Theory}, vol.~58,
  no.~5, pp. 2911--2934, May 2012.

\bibitem{mu_mimo}
Q.~H. Spencer, C.~B. Peel, A.~L. Swindlehurst, and M.~Haardt, ``An introduction
  to the multi-user {MIMO} downlink,'' \emph{{IEEE} Commun. Mag.}, vol.~42,
  no.~10, pp. 60--67, Oct. 2004.

\bibitem{boccardi_icassp07}
F.~Boccardi and H.~Huang, ``A near-optimum technique using linear precoding for
  the {MIMO} broadcast channel,'' in \emph{Proc. IEEE International Conference
  on Acoustics, Speech and Signal Processing (ICASSP)}, Honolulu (HI), Apr.
  2007.

\bibitem{set_packing_ref}
K.~Hoffman and M.~Padberg, ``Set covering, packing and partitioning problems,''
  \emph{Springer Encyclopedia of Optimization}, 2000.

\bibitem{lte_baker}
S.~Sesia, I.~Toufik, and M.~Baker, \emph{{LTE}: The {UMTS} {L}ong {T}erm
  {E}volution}.\hskip 1em plus 0.5em minus 0.4em\relax John Wiley \& Sons,
  2009.

\bibitem{wrap_ref}
T.~Hyt{\"{o}}nen, ``Optimal wrap-around network simulation,'' Helsinki
  University of Technology, Report A432, 2001.

\bibitem{mimo_kronecker_model}
J.~P. Kermoal, L.~Schumacher, K.~I. Pedersen, P.~E. Mogensen, and
  F.~Frederiksen, ``A stochastic {MIMO} radio channel model with experimental
  validation,'' \emph{{IEEE} J. Sel. Areas Commun.}, vol.~20, no.~6, pp.
  1211--1226, Aug. 2002.

\bibitem{tse_pfs}
P.~Viswanath, D.~Tse, and R.~Laroia, ``Opportunistic beamforming using dumb
  antennas,'' \emph{{IEEE} Trans. Inf. Theory}, vol.~48, no.~6, pp. 1277--1294,
  Jun. 2002.

\bibitem{benv_02}
N.~Benvenuto and G.~Cherubini, \emph{Algorithms for communications systems and
  their applications}.\hskip 1em plus 0.5em minus 0.4em\relax John Wiley \&
  Sons, 2002.

\bibitem{Rappaport}
T.~Rappaport, \emph{Wireless communications: principles and practice}.\hskip
  1em plus 0.5em minus 0.4em\relax Prentice Hall, 2002.

\end{thebibliography}

\end{document}